\documentclass[manuscript]{emulateapj}

\shorttitle{Self-cancellation of ephemeral regions}
\shortauthors{Yang et al.}

\slugcomment{Accepted for publication in ApJL}

\begin{document}

\title{Self-cancellation of ephemeral regions in the quiet Sun}

\author{Shuhong Yang\altaffilmark{1}, Jun Zhang\altaffilmark{1}, Ting Li\altaffilmark{1},
Yang Liu\altaffilmark{2}}

\altaffiltext{1}{Key Laboratory of Solar Activity, National
Astronomical Observatories, Chinese Academy of Sciences, Beijing
100012, China; [shuhongyang;zjun;liting]@nao.cas.cn}

\altaffiltext{2}{W.W. Hansen Experimental Physics Laboratory,
Stanford University, Stanford, CA 94305-4085, USA;
yliu@sun.stanford.edu}

\begin{abstract}

With the observations from the Helioseismic and Magnetic Imager
aboard the \emph{Solar Dynamics Observatory}, we statistically
investigate the ephemeral regions (ERs) in the quiet Sun. We find
that there are two types of ERs: normal ERs (NERs) and
self-cancelled ERs (SERs). Each NER emerges and grows with
separation of its opposite polarity patches which will cancel or
coalesce with other surrounding magnetic flux. Each SER also emerges
and grows and its dipolar patches separate at first, but a part of
magnetic flux of the SER will move together and cancel gradually,
which is described with the term ``self-cancellation" by us. We
identify 2988 ERs among which there are 190 SERs, about 6.4\% of the
ERs. The mean value of self-cancellation fraction of SERs is 62.5\%,
and the total self-cancelled flux of SERs is 9.8\% of the total ER
flux. Our results also reveal that the higher the ER magnetic flux
is, (i) the easier the performance of ER self-cancellation is, (ii)
the smaller the self-cancellation fraction is, and (iii) the more
the self-cancelled flux is. We think that the self-cancellation of
SERs is caused by the submergence of magnetic loops connecting the
dipolar patches, without magnetic energy release.

\end{abstract}

\keywords{Sun: activity --- Sun: photosphere --- Sun: surface
magnetism}

\section{Introduction}

Ephemeral regions (ERs) are short-lived small-scale dipolar magnetic
regions which were described by Harvey \& Martin (1973) for the
first time. According to the early results, ERs have a typical
lifetime of 1--2 days and a dimension of 30 000 km with a maximum
total flux of the order of 10$^{20}$ Mx (Harvey \& Martin 1973). In
the following years, the lifetime of ERs determined with higher
cadence observations was found to be much shorter, from around 12 hr
(Harvey et al. 1975) to less than 3 hr (Title 2000), and the
magnetic flux was found to be smaller ($\sim$ 10$^{19}$ Mx;
Schrijver et al. 1998; Chae et al. 2001).

In the quiet Sun, magnetic flux disappears due to cancellation and
dispersion, but at the same time it is continuously replenished by
newly emerged ERs. It takes only about one day for once replacement
of the flux in the quiet photosphere (Schrijver et al. 1998;
Hagenaar et al. 2003). Harvey et al. (1975) compared the parameters
of ERs and regular active regions and argued that ERs are the
small-scale end of a broad spectrum of magnetic activity. The
results from Martin (1988) and Hagenaar et al. (2003) reveal that
ERs likely vary in anti-phase with the solar cycle. The origin of
ERs is still not well known. The source may be the recycled flux
from decayed active regions (Nordlund 1992; Ploner et al. 2001).
Alternatively, they may be produced through local dynamo processes,
i.e., formed as a consequence of convective motions closer to the
surface (Hagenaar et al. 2003; Stein et al. 2003). Hagenaar et al.
(2008) studied the distribution and evolution of network magnetic
elements in the quiet Sun and found that the emergence rate of ERs
depends on the imbalance of magnetic flux surrounding the emergence
sites.

Yang et al. (2009) investigated magnetic field evolution in a
coronal hole region. They reported that an ER emerged and its
dipolar patches separated first and then moved together and
cancelled with each other. With the help of vector magnetic fields
and Doppler observations from the \emph{Hinode}, they concluded that
the cancellation between the opposite polarities of the ER was due
to the submergence of original loops that emerged from below the
photosphere. Recently, Wang et al. (2012) studied the solar
intranetwork magnetic elements in a quiet region and an enhanced
network area using the Narrow-band Filter Imager (NFI) magnetograms
from the \emph{Hinode}, and found an intranetwork dipolar flux
emergence followed by cancellation of its two poles with opposite
polarities. They believed that, after emergence, the dipolar flux
indeed submerged, i.e., retracted back into the sub-photosphere
again, because they had tracked the dipole continuously in the
magnetograms with high temporal and spatial resolutions.

Then one question is raised: in the quiet Sun, how often and to what
extent do ERs perform the behavior as reported by Yang et al. (2009)
and Wang et al. (2012), i.e., an ER cancelling itself after
emergence? The present paper is dedicated to answering this question
based on statistical results.

\section{Observations and data analysis}

The Helioseismic and Magnetic Imager (HMI; Scherrer et al. 2012;
Schou et al. 2012) on board the \emph{Solar Dynamics Observatory}
(\emph{SDO}; Pesnell et al. 2012) provides magnetic fields in the
full-disk of the Sun with a pixel size of 0$\arcsec$.5 and a cadence
of 45 s uninterruptedly. In this paper, we adopt the HMI full-disk
line-of-sight magnetograms with a 3 min cadence, i.e., one frame in
four, from 2010 June 11 12:00 UT to 2010 June 15 12:00 UT.

\begin{figure*}
\centering
\includegraphics
[bb=135 259 458 598,clip,angle=0,scale=0.85]{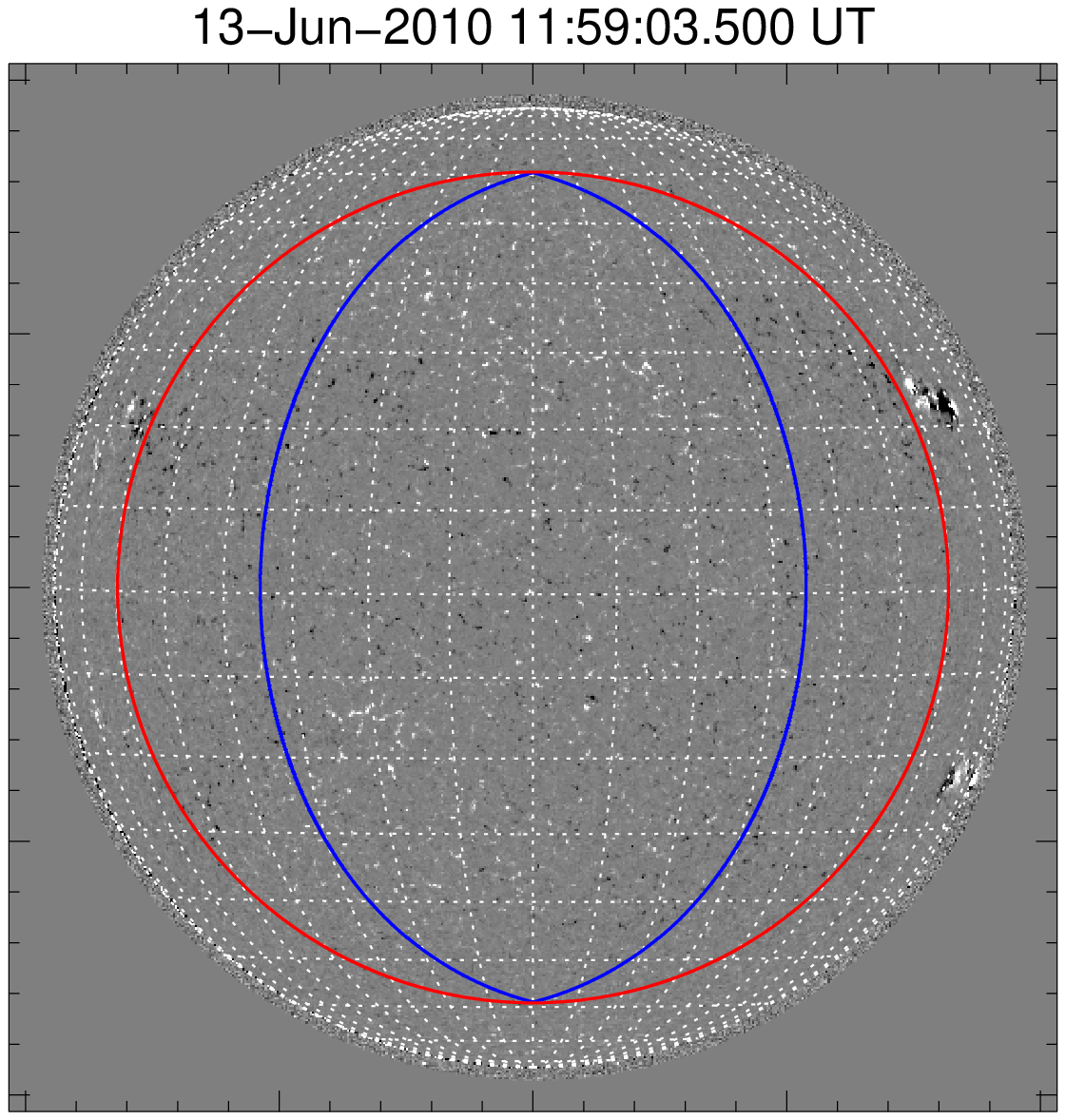} \caption{
{Line-of-sight magnetogram obtained by the \emph{SDO}}$/$HMI. The
red circle delineates the area where the heliocentric angles are
smaller than 60{\degr}. The blue curve outlines the pixels with
heliocentric angles smaller than 60{\degr} at all the times from
2010 June 11 12:00 UT to 2010 June 15 12:00 UT considering the
differential rotation of the Sun. \label{fig1}}
\end{figure*}

For each point ($x$, $y$) on the solar disk, the heliocentric angle
$\alpha$ is defined as: $\sin(\alpha) =
\sqrt{x^{2}+y^{2}}/R_{\odot}$, where $R_{\odot}$ is the solar radius
and the disk center is at $x=y=0$. Since the noise increases for
large $\alpha$, we do not consider the pixels with $\alpha$
$\geqslant$ 60{\degr} in each magnetogram as Hagenaar (2001) and Jin
et al. (2011) have done. In Figure 1, the red circle outlines the
area where $\alpha$ $<$ 60{\degr}. Then we derotate all the
magnetograms differentially to a reference time (2010 June 13 12:00
UT). Our target area is outlined with the blue curve within which
$\alpha$ $<$ 60{\degr} at all the times during the observation
period of our data set.

For each pixel in the rotated frame, there is a certain
$\alpha_{0}$. The pixel area \emph{S} in the HMI magnetograms is
0$\arcsec$.5 $\times$ 0$\arcsec$.5, and one pixel at $\alpha_{0}$
corresponds to a real area $S/\cos(\alpha_{0})$ on the solar
surface. The observed flux density $B$ is assumed to be related to
the flux density along the local normal direction, and then the
corrected flux density is $B/\cos(\alpha_{1})$, where $\alpha_{1}$,
instead of $\alpha_{0}$, is the real position at the observation
time.

\section{Results}

We track the ERs that emerged within the area outlined by the blue
curve in Figure 1, and identify 2988 ERs during the four days. We
find that these ERs can be classified into two types according to
their performance during evolution: normal ERs (NERs) and
self-cancelled ERs (SERs). For each type, we provide one movie
including three examples.

\subsection{Type one: NERs}

As can be seen from Movie 1 (available in the online edition), each
NER emerged and grew with separation of its opposite polarity
patches which cancelled or coalesced with other magnetic flux at the
end. Most of ERs are NERs, and the number of NERs is 2798.

The first example of NERs is shown in Figure 2. In the ellipse
region, the NER emerged and could be identified as a dipolar region
at 20:32 UT on June 13 (denoted by arrows in panel (a)). The two
patches with opposite polarities (labeled with ``A" and ``B") grew
and separated along the long axis (panel (b)). The negative patch
moved toward the pre-existing positive magnetic field (indicated by
arrow ``C") and cancelled with each other (panel (c)). At 02:05 UT
on June 14, most flux of ``A" had disappeared and the cancellation
between ``A" and ``C" was still going on (panel (d)).

\begin{figure*}
\centering
\includegraphics
[bb=65 302 503 594,clip,angle=0,scale=1.05]{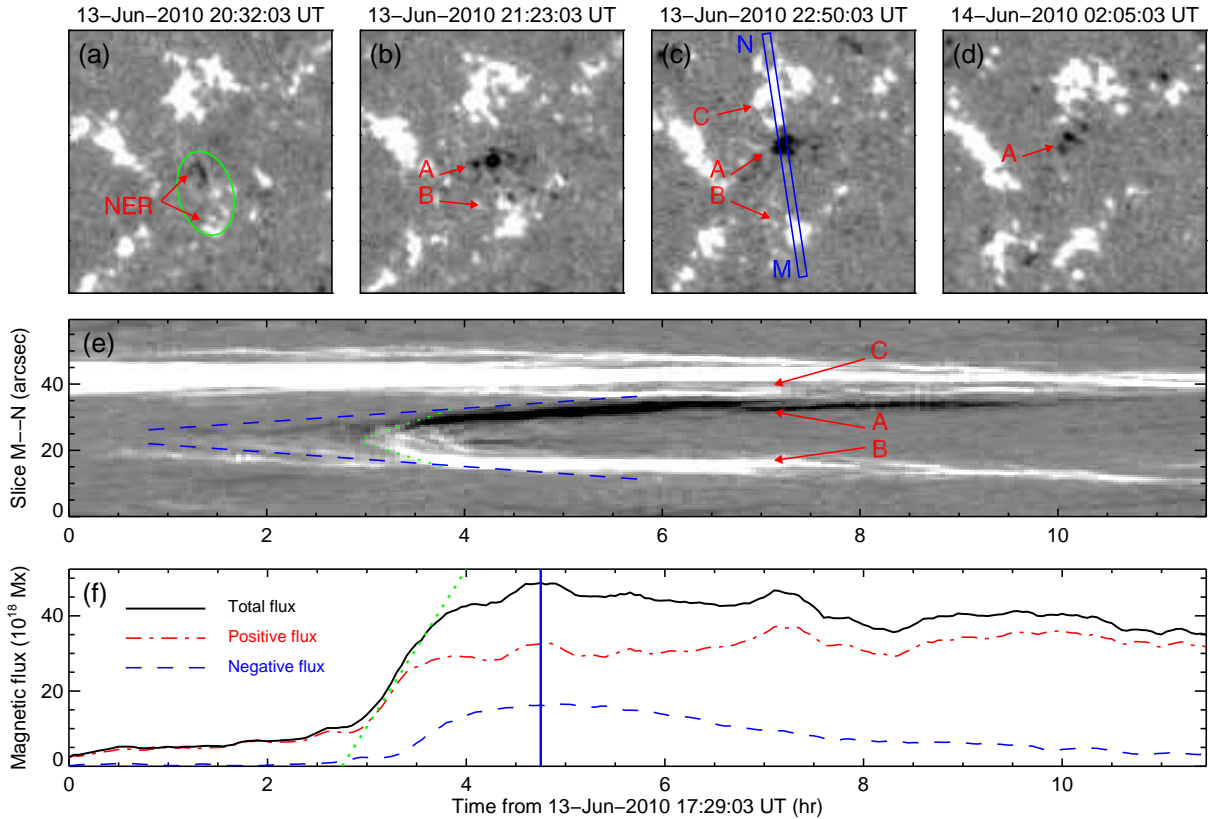} \caption{
\emph{Panels (a)--(d)}: sequence of HMI magnetograms displaying the
evolution of a NER denoted by the arrows and outlined by the ellipse
in panel (a). The field-of-view is 50{\arcsec} $\times$ 50{\arcsec}
and the grayscale saturates at $\pm$ 80 Mx cm$^{-2}$. Arrows ``A",
``B", and ``C" denote the negative and the positive patches of the
NER and the pre-existing positive magnetic fields, respectively.
\emph{Panel (e)}: space--time plot along slit ``M---N" marked in
panel (c). The dashed lines indicate the separation of the NER
patches, and the dotted lines a rapid emergence phase of the NER.
\emph{Panel (f)}: temporal variations of the positive (dash-dotted
curve), negative (dashed curve), and total (solid curve) magnetic
flux of the NER. The dotted line indicates a sharp increase of total
magnetic flux and the solid vertical line marks the flux maximum.
\label{fig2}}
\end{figure*}

Along slit ``M---N" marked in panel (c), we obtain the image profile
by averaging five pixels in the magnetograms in the direction
perpendicular to ``M---N". Then we make the space-time plot of such
profiles over time from June 13 17:29 UT to June 14 04:56 UT, as
displayed in panel (e). The NER emerged around 18:30 UT and the two
patches separated with an average velocity of 0.8 km s$^{-1}$ in the
first three hours (marked by the dashed lines). Moreover, during the
emerging stage, there was a rapid emergence with an expansion
velocity of 3.9 km s$^{-1}$ from 20:29 UT (marked by the dotted
lines). After 21:30 UT, when the NER was well developed, the
separation slowed down, and patch ``A" moved toward ``C" and
cancelled with it.

We measure the magnetic flux ($\phi$) of the NER by calculating
$\phi_{+}$ and $\phi_{-}$ in an area containing the two patches. The
selected area is changed according to the expansion of the NER to
ensure that the area has an appropriate size. After selection, the
pixels with unsigned magnetic fields weaker than 10.2 Mx cm$^{-2}$
(noise level determined by Liu et al. 2012) are eliminated. The
positive flux ($|\phi_{+}|$), negative flux ($|\phi_{-}|$), and
total flux ($|\phi_{+}|$+$|\phi_{-}|$) are plotted as a function of
time in panel (f). The total flux displayed a sharp increase at a
rate of 4.2 $\times$ 10$^{19}$ Mx hr$^{-1}$ during the rapid
emergence stage (20:29 UT --- 21:11 UT) and then appeared as a slow
increase and reached 4.8 $\times$ 10$^{19}$ Mx. From 22:20 UT when
the negative patch ``A" collided and cancelled with the pre-existing
positive patch ``C", the negative flux of the NER began to decrease
at a rate of 2.2 $\times$ 10$^{18}$ Mx hr$^{-1}$ , while the
positive flux of the NER mainly maintained at a steady level.

\subsection{Type two: SERs}

As shown in Movie 2 (available in the online edition), each SER
emerged and grew and its dipolar patches separated at first, but
then a part of magnetic flux of the SER moved together and cancelled
gradually, which is described with the term ``self-cancellation" by
us in this paper. There are only 190 SERs, about 6.4\% of the ERs.

Figure 3 shows the evolution of the first SER in Movie 2. The
ellipse region in panel (a) outlines the location of the SER. At
08:11 UT on June 14, the SER (indicated by arrows in panel (a)) had
exhibited a significant dipolar configuration. Then the SER went on
developing and its two patches (denoted by arrows ``A" and ``B" in
panel (b)) separated. Gradually, the positive polarity patch ``A"
split into two major elements ``A1" and ``A2", and ``A2" moved
toward to ``B" while ``A1" and ``B" did not change their positions a
lot (panel (c)). When ``A2" met with ``B", the cancellation took
place (see panel (d)), and at last, ``A2" completely disappeared and
part of ``B" was remained.

\begin{figure*}
\centering
\includegraphics
[bb=65 302 503 594,clip,angle=0,scale=1.05]{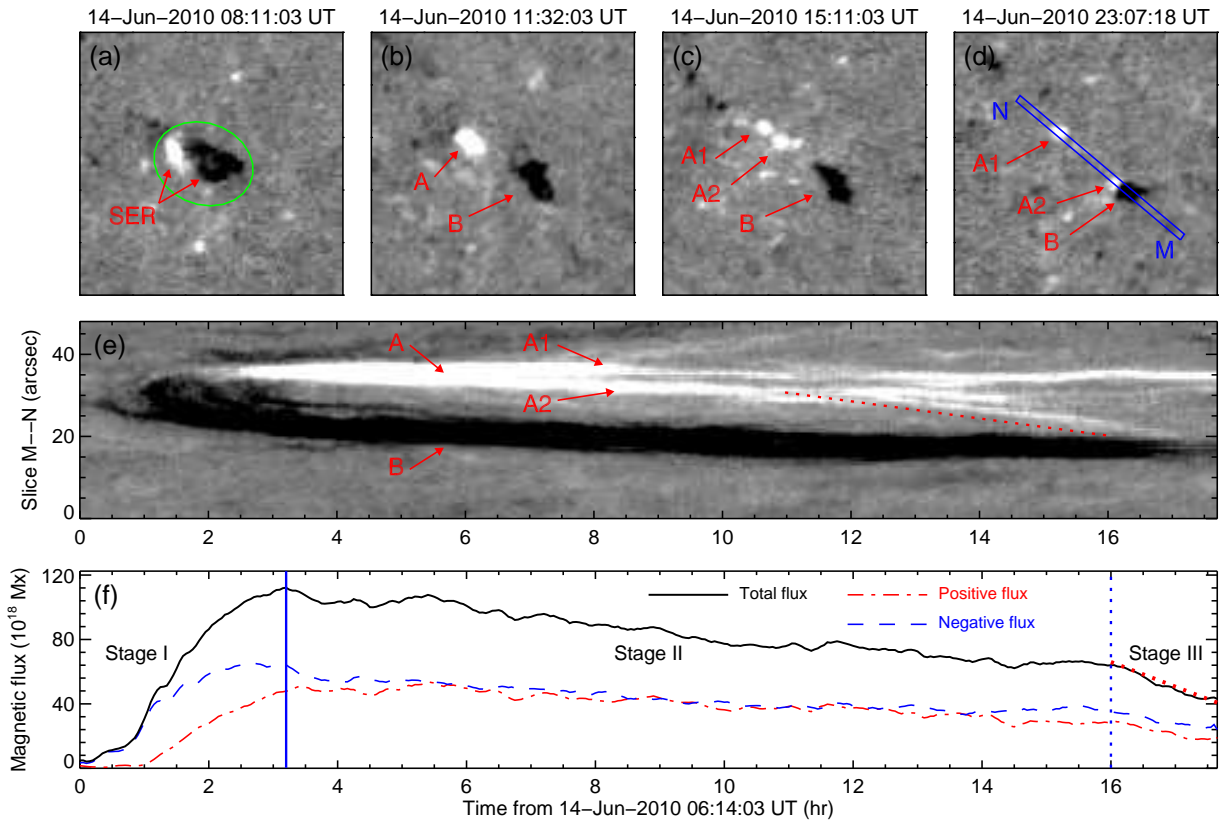} \caption{Similar
to Figure 2 but for the evolution of a SER. Magnetic elements ``A1"
and ``A2" are two components splitting from the positive patch of
the SER. The dotted line in panel (e) marks the movement of element
``A2". The red dotted line in panel (f) indicates the decrease of
total magnetic flux due to the cancellation between ``A2" and ``B",
and the dotted vertical line marks the start time of the
cancellation. \label{fig5}}
\end{figure*}

Along the separation and approach direction, i.e., ``M---N" marked
in panel (d), we obtain a series of image profiles as we have done
in Figure 2, and stack them in panel (e). The changes of positive
flux, negative flux, and total flux ($|\phi_{+}|$, $|\phi_{-}|$, and
$|\phi_{+}|$+$|\phi_{-}|$, respectively) during this period are
shown in panel (f). The SER violently emerged with the negative
patch followed by the positive one (the first three hours in panel
(e)), leading to a rapid increase of total flux to 1.1 $\times$
10$^{20}$ Mx and an imbalance between the positive and the negative
flux at the emergence stage (stage 1 in panel (f)). When the SER was
well developed, expansion of the two patches (positive ``A" and
negative ``B" in panel (e)) almost stopped, and patch ``A" began to
split into two components, ``A1" and ``A2". Component ``A2" moved
toward to ``B" (marked by the red dotted line in panel (e)) and met
with ``B" around 22:15 UT (marked by the dotted vertical line in
panel (f)). Then, ``A2" began to cancel with ``B", resulting in the
total disappearance of ``A2" and the shrinkage of ``B". This
self-cancellation process led to the significant decrease of total
magnetic flux from 6.4 $\times$ 10$^{19}$ Mx to 4.4 $\times$
10$^{19}$ Mx at a rate of 1.3 $\times$ 10$^{19}$ Mx hr$^{-1}$
(marked by the red dotted line in panel (f)).

\subsection{Statistical results of the ERs}

The probability density functions (PDFs) of the ER (black curve) and
SER (red curve) numbers are plotted in Figure 4(a). We find that the
PDF peak of the ERs is at 5.0 $\times$ 10$^{18}$ Mx, and there is
also a peak of the SER PDF which can be seen clearly in panel (b).
The PDF of SERs is peaked at 1.8 $\times$ 10$^{19}$ Mx. The blue
curve in panel (a) represents the number ratio of SER to ER. The
dash-dotted vertical line is located at 5.5 $\times$ 10$^{19}$ Mx, a
general separation of abundant samples (before) and few samples
(after). Thus, we think the ratio curve before the vertical line is
statistically reliable. The ratio curve increases from 0 to 0.5 with
the variation of ER flux from 0 to 5.5 $\times$ 10$^{19}$ Mx,
revealing that the higher the ER magnetic flux is, the easier the
performance of ER self-cancellation is.

\begin{figure*}
\centering
\includegraphics
[bb=43 150 512 666,clip,angle=0,scale=1.05]{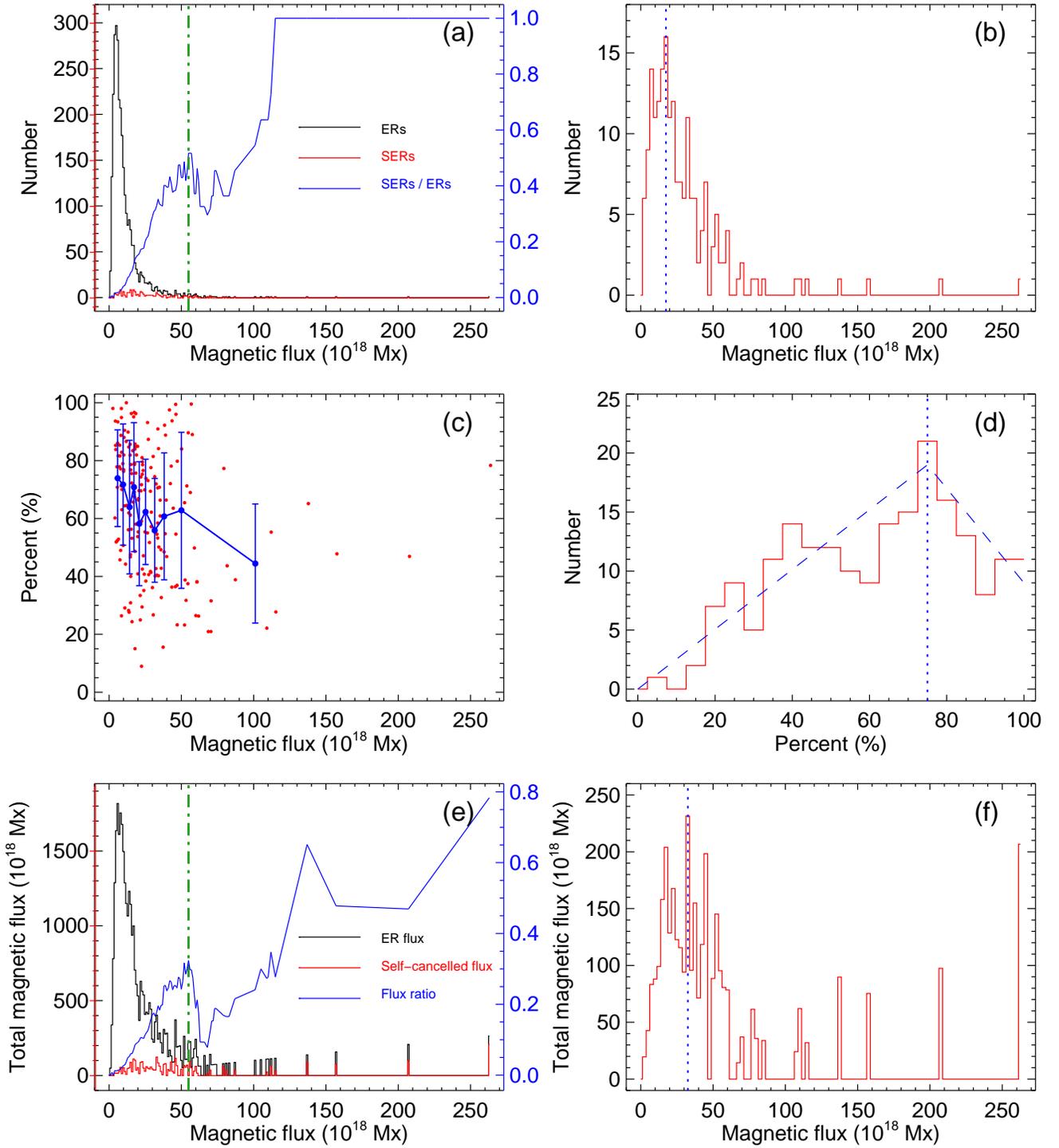} \caption{
\emph{Panel (a)}: PDFs of ER (black curve) and SER (red curve)
numbers and the number ratio (blue curve) of SER to ER. The binsize
is 1$\times$10$^{18}$ Mx. The dash-dotted vertical line is located
at 5.5 $\times$ 10$^{19}$ Mx, a general separation of abundant
samples (before) and few samples (after). \emph{Panel (b)}: PDF of
SERs at an enlarged displaying scale with a binsize of
2$\times$10$^{18}$ Mx. The dotted vertical line marks the PDF
maximum. \emph{Panel (c)}: scatter plots of self-cancellation
fractions versus magnetic flux of SERs (red symbols) and of sorted
and grouped points with error bars (blue symbols). \emph{Panel (d)}:
PDF of self-cancellation fractions with a binsize of 5\%. The dotted
vertical line marks the PDF maximum before and after which there are
increase and decrease trends (dashed lines). \emph{Panels (e) and
(f)}: similar to panels (a) and (b), but for ER flux (black curve),
self-cancelled flux (red curve), and the ratio (blue curve) of
self-cancelled flux to ER flux. \label{fig5}}
\end{figure*}

For SERs, only part of magnetic flux is self-cancelled. The scatter
plots of self-cancellation fractions versus magnetic flux of SERs
are shown with red symbols in Figure 4(c). The mean value of
self-cancellation fraction of SERs is 62.5\%. We apply a
``sort-group" method introduced by Zhao et al. (2009) to the SERs:
(1) all the SERs are sorted according to the total magnetic flux of
individuals; (2) the 190 sorted SERs are grouped into 10 data points
with equal SER number and each data point is assigned with the mean
value of the corresponding SER group; (3) the self-cancellation
fractions and the magnetic flux values are correlated with each
other and plotted with blue symbols in Figure 4(c). The statistical
correlation shows a general tendency that the higher the ER magnetic
flux is, the smaller the self-cancellation fraction is. We plot the
PDF of self-cancellation fractions in panel (d). The dotted vertical
line marks the PDF maximum at 75\%. We can see that there are an
increase and a decrease trends before and after the PDF peak,
respectively.

Figures 4(e) and 4(f) are similar to Figures 4(a) and 4(b), but for
ER flux (black curve), self-cancelled flux (red curve), and the
ratio (blue curve) of self-cancelled flux to ER flux. The PDF of ER
flux is peaked at 7.0 $\times$ 10$^{18}$ Mx (panel (e)), and the
peak of self-cancelled flux is at 3.3 $\times$ 10$^{19}$ Mx (panel
(f)). The flux ratio curve exhibits a variation from 0 to 0.3 in the
range of 0 --- 5.5 $\times$ 10$^{19}$ Mx, indicating that the higher
the ER magnetic flux is, the more the self-cancelled flux is.

\section{Conclusions and discussion}

With the observations from the \emph{SDO}/HMI, we statistically
investigate the ERs in the quiet Sun. We find that there are two
types of ERs: NERs and SERs. Each NER emerged and grew with
separation of its opposite polarity patches which finally cancelled
or coalesced with other magnetic flux. Each SER also emerged and
grew and its dipolar patches separated at first, but then a part of
magnetic flux of the SER moved together and cancelled gradually,
which is described with the term ``self-cancellation". We identify
2988 ERs among which there are 190 SERs, about 6.4\% of the ERs. The
mean value of self-cancellation fraction of SERs is 62.5\%, and the
total self-cancelled flux of SERs is 9.8\% of the total ER flux. Our
results also reveal that the higher the ER magnetic flux is, (i) the
easier the performance of ER self-cancellation is, (ii) the smaller
the self-cancellation fraction is, and (iii) the more the
self-cancelled flux is.

Magnetic flux cancellation is an observational phenomenon of flux
disappearance when two magnetic patches with different polarities
encounter (Livi et al. 1985; Martin et al. 1985; Zhang et al. 2001).
As one of the three modes for removal of magnetic flux with opposite
polarities from the photosphere illustrated by Zwaan (1978, 1987),
the disappearance of magnetic flux can be resulted from the
retraction of magnetic loops into the sub-photosphere, if the two
poles are still connected by initial loops. So in theory, the
process that magnetic loops emerge into the solar atmosphere and
then submerge below the subsurface is quite reasonable and possible.
The submergence of part of an active region was studied by Rabin et
al. (1984) and the submergence of a sunspot group was reported by
Zirin (1985) who suggested that submergence of an active region may
be common. To determine if cancellation is caused by the submergence
of magnetic loops, it is important to check vector field data. With
\emph{Hinode }spectro-polarimetric data, Yang et al. (2009) found
that at the area where the two opposite polarities of the ER
cancelled, there were strong transverse fields pointing directly
from the positive patch to the negative one. Moreover, they also
observed larger Doppler redshifts between the cancelling patches.
They suggested that the cancellation of the ER was due to the
submergence of original loops. In the recent study of Wang et al.
(2012), although there was a lack of vector field observations, they
believed that the emergence and submergence of the ER they observed
was a real behavior since the ER had been tracked in the high
tempo-spatial resolution photospheric magnetogram. Thus, we also
think the self-cancellation of SERs is caused by the submergence of
magnetic loops connecting the dipolar patches.

When magnetic patches of ERs separate and cancel with other magnetic
elements, the connection between opposite polarities will be changed
and the magnetic loops will be restructured to a lower potential
configuration, which requires magnetic flux reconnection accompanied
with energy release (Wang \& Shi 1993). While when the ER loops
emerge through the photosphere layer from below and then submerge
into the sub-photosphere again, no magnetic reconnection occurs and
no magnetic energy is released.

Our results reveal a tendency that the higher the ER magnetic flux
is, the easier the performance of ER self-cancellation is. We
suggest that the behaviour depends on the magnitude of magnetic flux
and on the relative importance and balance between the magnetic
pressure and the magnetic tension acting on the emerging flux tubes.
In this study, we also notice that the dipolar patches of most of
the SERs split before the self-cancellation phase (eg., as shown in
Figure 3). It may be caused by plasma motions in the photosphere.
When they emerge into the photosphere, they are drifted toward the
supergranular boundaries by mesogranular and supergranular flow and
split into smaller fragments due to granular convection (Simon et
al. 2001; Priest et al. 2002).

\acknowledgments { We are grateful to the referee for helpful
comments. We thank Prof. Jingxiu Wang for his valuable suggestions.
This work is supported by the National Natural Science Foundations
of China (40890161, 11025315, 10921303, 41074123 and 11003024), the
CAS Project KJCX2-EW-T07, the National Basic Research Program of
China under grant 2011CB811403, and the Young Researcher Grant of
National Astronomical Observatories, Chinese Academy of Sciences.
The data have been used by courtesy of NASA/\emph{SDO} and the HMI
science team.}

{}



\begin{thebibliography}{}

\bibitem[Chae et al.(2001)]{2001ApJ...548..497C}
Chae, J., Martin, S.~F., Yun, H.~S., et al.\ 2001, \apj, 548, 497
\bibitem[Hagenaar(2001)]{2001ApJ...555..448H}
Hagenaar, H.~J.\ 2001, \apj, 555, 448
\bibitem[Hagenaar et al.(2008)]{2008ApJ...678..541H}
Hagenaar, H.~J., De Rosa, M.~L., \& Schrijver, C.~J.\ 2008, \apj,
678, 541
\bibitem[Hagenaar et al.(2003)]{2003ApJ...584.1107H}
Hagenaar, H.~J., Schrijver, C.~J., \& Title, A.~M.\ 2003, \apj, 584,
1107
\bibitem[Harvey et al.(1975)]{1975SoPh...40...87H}
Harvey, K.~L., Harvey, J.~W., \& Martin, S.~F.\ 1975, \solphys, 40,
87
\bibitem[Harvey \& Martin(1973)]{1973SoPh...32..389H}
Harvey, K.~L., \& Martin, S.~F.\ 1973, \solphys, 32, 389
\bibitem[Jin et al.(2011)]{2011ApJ...731...37J}
Jin, C.~L., Wang, J.~X., Song, Q., \& Zhao, H.\ 2011, \apj, 731, 37
\bibitem[()]{}
Liu, Y., Hoeksema, J. T., Scherrer, P. H., et al. 2012, \solphys, in
press.
\bibitem[Livi et al.(1985)]{1985AuJPh..38..855L}
Livi, S.~H.~B., Wang, J., \& Martin, S.~F.\ 1985, Australian Journal
of Physics, 38, 855
\bibitem[Martin(1988)]{1988SoPh..117..243M}
Martin, S.~F.\ 1988, \solphys, 117, 243
\bibitem[Martin et al.(1985)]{1985AuJPh..38..929M}
Martin, S.~F., Livi, S.~H.~B., \& Wang, J.\ 1985, Australian Journal
of Physics, 38, 929
\bibitem[Nordlund et al.(1992)]{1992ApJ...392..647N}
Nordlund, A., Brandenburg, A., Jennings, R.~L., et al.\ 1992, \apj,
392, 647
\bibitem[Pesnell et al.(2012)]{2012SoPh..275....3P}
Pesnell, W.~D., Thompson, B.~J., \& Chamberlin, P.~C.\ 2012,
\solphys, 275, 3
\bibitem[Ploner et al.(2001)]{2001ASPC..236..363P}
Ploner, S.~R.~O., Sch{\"u}ssler, M., Solanki, S.~K., \& Gadun,
A.~S.\ 2001, Advanced Solar Polarimetry -- Theory, Observation, and
Instrumentation, 236, 363
\bibitem[Priest et al.(2002)]{2002ApJ...576..533P}
Priest, E.~R., Heyvaerts, J.~F., \& Title, A.~M.\ 2002, \apj, 576,
533
\bibitem[Rabin et al.(1984)]{1984ApJ...287..404R}
Rabin, D., Moore, R., \& Hagyard, M.~J.\ 1984, \apj, 287, 404
\bibitem[Schou et al.(2012)]{2012SoPh..275..229S}
Schou, J., Scherrer, P.~H., Bush, R.~I., et al.\ 2012, \solphys,
275, 229
\bibitem[Scherrer et al.(2012)]{2012SoPh..275..207S}
Scherrer, P.~H., Schou, J., Bush, R.~I., et al.\ 2012, \solphys,
275, 207
\bibitem[Schrijver et al.(1998)]{1998Natur.394..152S}
Schrijver, C.~J., Title, A.~M., Harvey, K.~L., et al.\ 1998, \nat,
394, 152
\bibitem[Simon et al.(2001)]{2001ApJ...561..427S}
Simon, G.~W., Title, A.~M., \& Weiss, N.~O.\ 2001, \apj, 561, 427
\bibitem[Stein et al.(2003)]{2003ASPC..286..121S}
Stein, R.~F., Bercik, D., \& Nordlund, {\AA}.\ 2003, Current
Theoretical Models and Future High Resolution Solar Observations:
Preparing for ATST, 286, 121
\bibitem[Title(2000)]{2000RSPTA.358..657T}
Title, A.\ 2000, Royal Society of London Transactions Series A, 358,
657
\bibitem[Wang \& Shi(1993)]{1993SoPh..143..119W}
Wang, J. X., \& Shi, Z. X. \ 1993, \solphys, 143, 119
\bibitem[Wang et al.(2012)]{2012SoPh..tmp...49W}
Wang, J. X., Zhou, G. P., Jin, C. L., \& Li, H.\ 2012, \solphys,
278, 299.
\bibitem[Yang et al.(2009)]{2009ApJ...703.1012Y}
Yang, S.~H., Zhang, J., \& Borrero, J.~M.\ 2009, \apj, 703, 1012
\bibitem[Zhang et al.(2001)]{2001ApJ...548L..99Z}
Zhang, J., Wang, J. X., Deng, Y. Y., \& Wu, D. J.\ 2001, \apjl, 548,
L99
\bibitem[Zhao et al.(2009)]{2009RAA.....9..933Z}
Zhao, M., Wang, J.~X., Jin, C.~L., \& Zhou, G.~P.\ 2009, RAA, 9, 933
\bibitem[Zirin(1985)]{1985ApJ...291..858Z}
Zirin, H.\ 1985, \apj, 291, 858
\bibitem[Zwaan(1978)]{1978SoPh...60..213Z}
Zwaan, C.\ 1978, \solphys, 60, 213
\bibitem[Zwaan(1987)]{1987ARA&A..25...83Z}
Zwaan, C.\ 1987, \araa, 25, 83

\end{thebibliography}
\end{document}